# Block Scrambling Image Encryption Used in Combination with Data Augmentation for Privacy-Preserving DNNs

Tatsuya CHUMAN, *Student member*, *IEEE,* and Hitoshi KIYA, *Fellow, IEEE*
Department of Computer Science, Tokyo Metropolitan University, Tokyo, Japan

*Abstract*—In this paper, we propose a novel learnable image encryption method for privacy-preserving deep neural networks (DNNs). The proposed method is carried out on the basis of block scrambling used in combination with data augmentation techniques such as random cropping, horizontal flip and grid mask. The use of block scrambling enhances robustness against various attacks, and in contrast, the combination with data augmentation enables us to maintain a high classification accuracy even when using encrypted images. In an image classification experiment, the proposed method is demonstrated to be effective in privacy-preserving DNNs.

## I. INTRODUCTION

The spread use of deep neural networks (DNNs) has greatly contributed to solving complex tasks for many applications [1], including privacy-sensitive/security-critical ones such as facial recognition and medical image analysis. Recently, it has been very popular for data owners to utilize cloud servers to compute and process a large amount of data instead of using local servers. However, there are risks of data leakage in cloud environments [2]. Because application users (i.e. clients) want to avoid the risks, privacy-preserving DNNs have become an urgent challenge.

Various perceptual encryption methods have been proposed to generate visually-protected images [3]-[7]. Although information theory-based encryption (like RSA and AES) generates a ciphertext, images encrypted by the perceptual encryption methods can be directly applied to some image processing algorithms or image compression algorithms. Numerous encryption schemes have been proposed for privacy-preserving DNNs [5]-[7], but several attacks including DNN-based ones were shown to restore visual information from encrypted images [8]. Therefore, encryption schemes that are robust against various attacks are essential for privacy-preserving DNNs.

In this paper, we propose a block scrambling encryption used in combination with data augmentation techniques for privacy-preserving DNNs. Block scrambling is well-known to enhance robustness against attacks [9], but it decreases the performance of DNN modes. Accordingly, in this paper, a combination with data augmentation techniques is considered to improve this issue for the first time. The proposed encryption enables us to not only apply images without visual information to DNNs for both training and testing but to also consider the use of independent encryption keys, which means that all images are encrypted by using different security keys. In an image classification experiment, it is confirmed that the proposed method improves a classification accuracy by using a combination with data augmentation techniques.

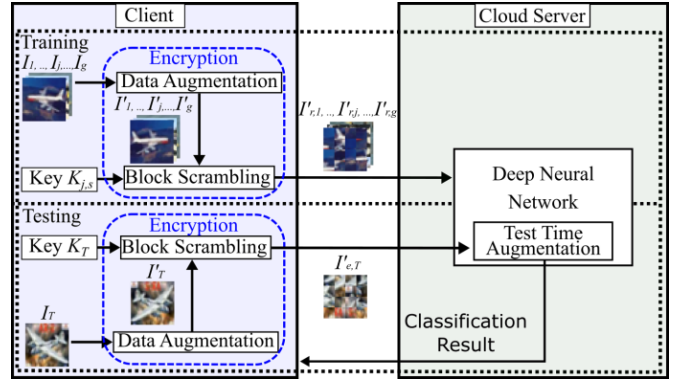

Fig. 1. Scenario of privacy-preserving deep neural networks

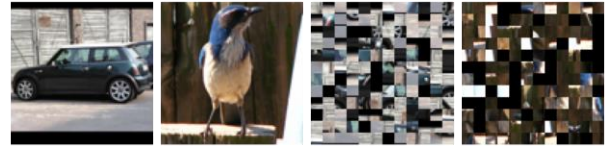

(a) Original images    (b) Encrypted images
Fig. 2. Example of encrypted images ($Bx = By = 8$)

## II. PROPOSED SCHEME

In this section, we propose a novel block scrambling method used in combination with data augmentation for privacy preserving DNNs. Figure 1 indicates the scenario of this paper.

*Encryption for training image*

The procedure of encryption for training images is given as follows.
1) A client applies data augmentation techniques to training images ($I_1, .., I_j, ..., I_g$) to generate modified ones ($I'_1, .., I'_j, ..., I'_g$), where $g$ is the number of training images, where random cropping, random horizontal flip and Gridmask [10] are carried out as data augmentation.
2) Every $I'_j$ is divided into $Bx \times By$ blocks, and then permute randomly the blocks to generate a permuted image from each $I'_j$ by using a secret key $K_{j,s}$, $s \in \{1, 2, ..., N\}$, where $s$ is randomly chosen per each epoch.
3) The permuted images are sent to a cloud server as encrypted images ($I'_{r,1}, .., I'_{r,j}, ..., I'_{r,g}$).
4) A model is trained by using the encrypted images.

*Encryption for testing image*

The procedure of encryption for a testing image $I_T$ is described as follows.
1) A client applies data augmentation to $I_T$ to generate modified one $I'_T$, where random cropping and random horizontal flip are carried out as data augmentation.

2) $I'_T$ is divided into $Bx \times By$ blocks, and then permute randomly the blocks to generate a permuted image by using a secret key $K_T$.
3) The permuted image is sent to a cloud server as an encrypted image $I'_{e,T}$.

*Classification with encrypted image*

1) Test time augmentation is applied to $I'_{e,T}$ in the cloud server. In this paper, horizontal flip is carried out as test time augmentation.
2) The cloud server returns a classification result obtained from the trained model to the client.

An example of encrypted images by using the proposed scheme ($Bx = By = 8$) is shown in Fig. 2(b); Fig. 2(a) shows the original ones. As illustrated in Fig.2(b), visual information on plain images is heavily decreased, thus the privacy of images is protected even if the cloud server leaks encrypted images.

Although jigsaw puzzle solver attacks have been proposed as cipher-text only attacks, it was confirmed that the use of small block enhances security against several attacks [3], [9]. In this paper, $Bx = By = 8$ is used as a block size. In addition, the combination with data augmentation techniques enhances difficulty in restoring the visual information.

III. SIMULATION

*Simulation condition*

We trained a CNN model for image classification by using the CIFAR-10 dataset, which contains of 50,000 training images and 10,000 testing images with 10 classes, and each image has 32×32 pixels. The model with ResNet-18 was trained for 250 epochs by using the stochastic gradient descent (SGD) optimizer under a batch size of 128, a momentum of 0.9, weight decay of 0.0005, and maximum learning rate of 0.0005. As we mentioned above, random cropping with a padding of 4, random horizontal flip and Gridmask were carried out as data augmentation techniques for the training images. In contrast, random cropping with a padding of 4 and random horizontal flip were carried out for the testing images. Furthermore, 4 permutated images were generated from every modified image, namely $N = 4$ was selected in the simulations.

*Simulation result*

Figure 3 show experiment results under two conditions: no data augmentation, data augmentation and test time augmentation. In Fig.3 (a), only block scrambling was applied. In contrast, data augmentation techniques were applied in addition to block scrambling, and moreover test time augmentation was carried out in Fig.3 (b).

As shown in Fig. 3(b), the proposed method was confirmed to improve a classification performance under the use of not only plain images but also encrypted ones, where the difference of accuracies between plain images and encrypted ones was reduced compared to the difference in Fig.3(a). Note that block scrambling used in combination with data augmentation also has a higher difficulty in restoring visual information on plain images than simple block scrambling.

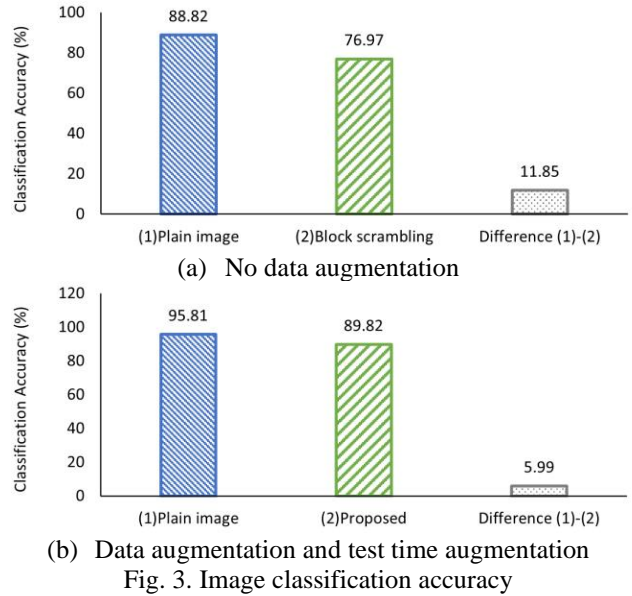

(a) No data augmentation

(b) Data augmentation and test time augmentation

Fig. 3. Image classification accuracy

IV. CONCLUSION

In the paper, we proposed a novel block scrambling encryption scheme used in combination with data augmentation for privacy preserving DNNs. Block scrambling is well known to enhance robustness against various attacks, but it reduces the performance of a DNN model in general. In this paper, to maintain a high performance, a combination with data augmentation techniques was considered for the first time. In an image classification experiment, the proposed method was confirmed to reduce influence of using encrypted images on classification accuracy.